\documentclass[sigconf]{acmart}

\AtBeginDocument{%
  }

\setcopyright{none}

\acmDOI{}
\acmYear{}
\acmConference{}{}

\acmConference[WebSci’25]{Make sure to enter the correct
  conference title from your rights confirmation emai}{May 20--24,
  2025}{New Brunswick, NJ}

  \acmISBN{}

\usepackage{tabularx}
\usepackage[normalem]{ulem}
\usepackage{float}
\usepackage{stfloats} 

\begin{document}

\title{Analyzing Political Discourse on Discord during the\\2024 U.S. Presidential Election}



\author{Arthur Buzelin}
\email{arthurbuzelin@dcc.ufmg.br}
\authornote{These authors contributed equally to this research.}
\orcid{0009-0007-0816-7189}
\affiliation{%
  \institution{Universidade Federal de Minas Gerais}
  \city{Belo Horizonte}
  \country{Brazil}
}

\author{Pedro Robles Dutenhefner}
\email{pedroroblesduten@ufmg.br}
\authornotemark[1]
\orcid{1234-5678-9012}
\affiliation{%
  \institution{Universidade Federal de Minas Gerais}
  \city{Belo Horizonte}
  \country{Brazil}
}

\author{Marcelo Sartori Locatelli}
\email{locatellimarcelo@dcc.ufmg.br}
\authornotemark[1]
\orcid{0000-0002-0893-1446}
\affiliation{%
  \institution{Universidade Federal de Minas Gerais}
  \city{Belo Horizonte}
  \country{Brazil}
}

\author{Samira Malaquias}
\email{samiramalaquias@dcc.ufmg.br}
\orcid{1234-5678-9012}
\affiliation{%
  \institution{Universidade Federal de Minas Gerais}
  \city{Belo Horizonte}
  \country{Brazil}
}
\author{Pedro Bento}
\email{pedro.bento@dcc.ufmg.br}
\orcid{1234-5678-9012}
\affiliation{%
  \institution{Universidade Federal de Minas Gerais}
  \city{Belo Horizonte}
  \country{Brazil}
}

\author{Yan Aquino}
\email{yanaquino@dcc.ufmg.br}
\orcid{0009-0009-1647-4298}
\affiliation{%
  \institution{Universidade Federal de Minas Gerais}
  \city{Belo Horizonte}
  \country{Brazil}
}

\author{ Lucas Dayrell}
\email{lucasdayrell@dcc.ufmg.br}
\orcid{1234-5678-9012}
\affiliation{%
  \institution{Universidade Federal de Minas Gerais}
  \city{Belo Horizonte}
  \country{Brazil}
}

\author{Victoria Estanislau}
\email{victoria.estanislau@dcc.ufmg.br}
\orcid{1234-5678-9012}
\affiliation{%
  \institution{Universidade Federal de Minas Gerais}
  \city{Belo Horizonte}
  \country{Brazil}
}

\author{Caio Santana}
\email{caiosantana@dcc.ufmg.br}
\orcid{1234-5678-9012}
\affiliation{%
  \institution{Universidade Federal de Minas Gerais}
  \city{Belo Horizonte}
  \country{Brazil}
}

\author{Pedro Alzamora}
\email{pedro.loures@dcc.ufmg.br}
\orcid{1234-5678-9012}
\affiliation{%
  \institution{Universidade Federal de Minas Gerais}
  \city{Belo Horizonte}
  \country{Brazil}
}

\author{Marisa Vasconcelos}
\email{marisa.vasconcelos@gmail.com}
\orcid{1234-5678-9012}
\affiliation{%
  \institution{Universidade Federal de Minas Gerais}
  \city{Belo Horizonte}
  \country{Brazil}
}

\author{Wagner Meira Jr.}
\email{meira@dcc.ufmg.br}
\orcid{0000-0002-2614-2723}
\affiliation{%
  \institution{Universidade Federal de Minas Gerais}
  \city{Belo Horizonte}
  \country{Brazil}
}

\author{Virgilio Almeida}
\email{virgilio@dcc.ufmg.br}
\orcid{0000-0001-6452-0361 }
\affiliation{%
  \institution{Universidade Federal de Minas Gerais}
  \city{Belo Horizonte}
  \country{Brazil}
}

\renewcommand{\shortauthors}{Buzelin et al.}

\begin{abstract}
Social media networks have amplified the reach of social and political movements, but most research focuses on mainstream platforms such as X, Reddit, and Facebook, overlooking Discord. As a rapidly growing, community-driven platform with optional decentralized moderation, Discord offers unique opportunities to study political discourse. This study analyzes over 30 million messages from political servers on Discord discussing the 2024 U.S. elections. Servers were classified as Republican-aligned, Democratic-aligned, or unaligned based on their descriptions. We tracked changes in political conversation during key campaign events and identified distinct political valence and implicit biases in semantic association through embedding analysis. We observed that Republican servers emphasized economic policies and Democratic servers focusing on equality-related and progressive causes. Furthermore, we detected an increase in toxic language, such as sexism, in Republican-aligned servers after Kamala Harris's nomination. These findings provide a first look at political behavior on Discord, highlighting its growing role in shaping and understanding online political engagement.
\end{abstract}




\maketitle

\section{Introduction}
\label{sec:introduction}

Social networks have become a powerful tool for amplifying social movements, such as protests \cite{wolfsfeld2013social} and political campaigns~\cite{groshek2017helping}, allowing messages to reach audiences as large as those of traditional media like television -- all without relying on mainstream outlets~\cite{boynton2016agenda}.   
Due to these reasons, researchers have identified the important role of social media in political success. Indeed, during the 2008 U.S. presidential elections,  Barack Obama's use of social media as part of his campaign strategy was already regarded as a key factor in his victory~\cite{hughes2010obama}.

Since then, social media's influence on politics has only intensified. The spread of fake news, misinformation and bot-driven manipulation during the 2016 and 2020 elections~\cite{gunther2019fake,ferrara2020characterizing}, raised significant questions about the need for regulation and moderation on these platforms. As a result, platforms such as Facebook and X (formerly Twitter)~\cite{zannettou2021won} implemented interventions to curb the potential harm caused by these strategies. Despite these efforts, social media continues to enable harmful activities, such as spreading  conspiracy theories and organizing events like the January 2021 Capitol riot~\cite{o2023coming}. This highlights the importance of studying political discussions on social media platforms. 


Most research so far has focused on mainstream platforms, such as YouTube, X, Facebook, Instagram, and Reddit. However, other platforms may hold significant influence in shaping voters' opinions during the electoral process, still remaining under-explored. 
One of these platforms is Discord, which has experienced significant growth since the pandemic. Originally designed for gaming and related topics, it has since been adopted by many other communities~\cite{johnson2022embracing}.
The platform is structured into user-created and managed groups, which are called servers, where they can share text, images, videos, and voice-chat. 
Unlike mainstream social media platforms, moderation and guidelines on Discord are primarily determined by server creators \footnote{https://discord.com/community-moderation-safety}, meaning they may vary across different servers. This decentralized structure fosters less-moderated and unfiltered discussions, including those in public community servers.

Discord's unique characteristics make it a compelling subject for study. Its semi-private servers are similar to platforms like Telegram, but its recent controversies -- such as child abuse networks uncovered in Brazil~\cite{discord2024scandalbrazil} and its role in mental health discussions \cite{webmedia} -- show both its potential and risks. The growing presence of alt-right extremism on the platform~\cite{gallagher2021extreme} further highlights the importance of studying its impact on political discussions. Despite these emerging concerns, there remains a significant gap in scientific research regarding the political dimensions of this network, particularly  its role in influencing political discourse and the presence of already established ideological communities.
    
To the best of our knowledge, this study represents the first in-depth political analysis of Discord, focusing on the 2024 United States presidential election. Specifically, we aim to explore how the platform's unique characteristics shape political discussions and user interactions. We address the following research questions:

\begin{enumerate}
    \item RQ1: How does the discourse on Discord servers vary across different political spectra and electoral periods, particularly in response to major political events? 
    \item RQ2: What is the level of toxicity in political discussions on Discord, and which groups are the primary targets of hate speech across different political spectra and key electoral moments?
\end{enumerate}

We find that (1) by tracking shifts in political conversations during key campaign events, we identified distinct political valences and implicit biases in semantic associations through embedding analysis. Republican-aligned servers emphasized economic policies, while Democratic-aligned servers focused on equality-related and progressive causes. The volume of political messages surged significantly during pivotal events such as Biden's exit from the race and the presidential debates.


Additionally, we note that (2) the discussion on Republican-aligned servers was considerably more toxic than on Democratic-aligned servers, especially in the period just after Kamala Harris was announced as the Democrat candidate, which was correlated with increased toxicity and sexism on the Republican servers.

\begin{table*}[!h]
\small
\centering
\caption{Statistics of user and bot posts across server categories. Most of the messages belong to unaligned servers. The unique users and unique bots columns do not sum to total, as some users participate in discussion on multiple server categories.}
\begin{tabular}{crrrrr} \toprule
\textbf{Alignment} & \textbf{Unique Users} & \textbf{User Messages} & \textbf{Messages/User} & \textbf{Unique Bots} & \textbf{Bot messages} \\ \midrule
\textbf{ Democratic}    & 10,149                & 520,614                & 51.29                  & 69                   & 47,123                \\
\textbf{Unaligned} & 71,686                & 31,603,712             & 440.86                 & 618                  & 2,102,070             \\
\textbf{ Republican}   & 4,255                 & 555,750                & 130.61                 & 35                   & 76,412                \\
\textbf{Total}   & 83,611                & 32,680,076             & 390.85                 & 675                  & 2,225,605             \\ \bottomrule
\end{tabular}

\label{tab:statistics}
\end{table*}

\section{Related Work}
\label{sec:related_work}

\begin{figure*}[!ht]
    \centering
    \includegraphics[width=0.8\linewidth]{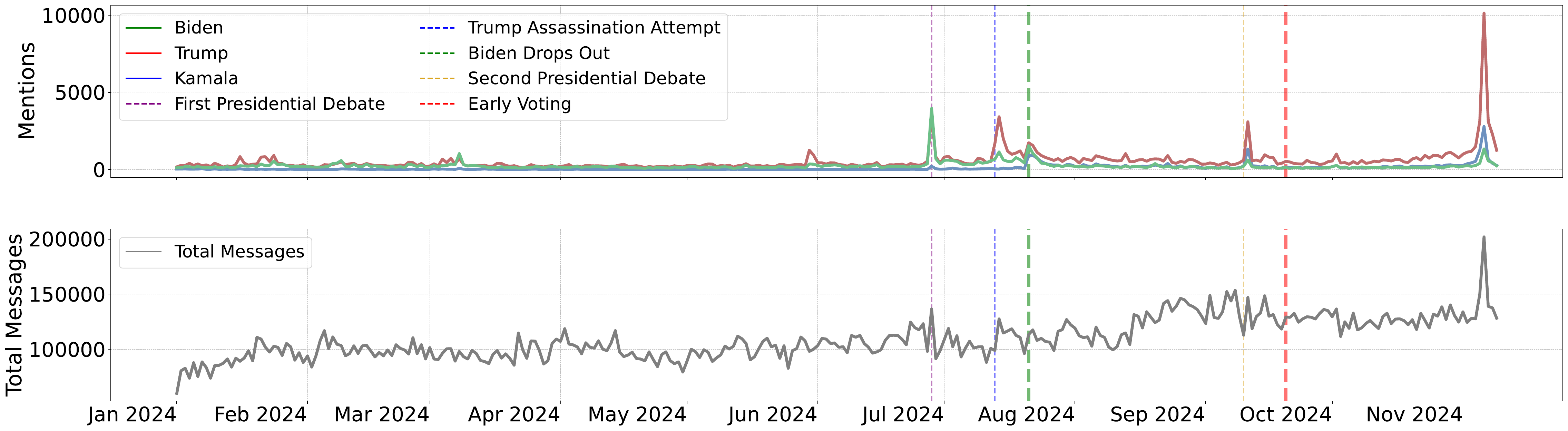}
    \caption{Volume of messages shared on Discord political servers, categorized by metadata. The first plot shows total daily messages across all servers. The second plot shows daily mentions of ``Trump'', ``Biden'', and ``Kamala''.}
    \label{fig:line-plot}
    \Description{Volume of messages shared on Discord political servers categorized by metadata. The first plot shows the total number of messages shared per day across all servers. The second plot depicts daily mentions of ``Trump'', ``Biden'', and ``Kamala''.}
\end{figure*}

The influence of social media on political engagement has been widely discussed, particularly in the context of mainstream platforms such as X and Facebook. For example, Fujiwara et al.~\cite{fujiwara2024effect} examined the role of Twitter in the 2016 and 2020 U.S. presidential elections and found that increased Twitter usage in certain counties contributed to a significant reduction in the Republican vote share. In addition, Allcott et al.~\cite{allcott24theeffects} investigated the effects of deactivating Facebook and Instagram accounts during the 2020 U.S. election. Their study revealed that while Instagram had no measurable impact on political outcomes, Facebook significantly increased user awareness of general news but also led to greater exposure to misinformation.

In more recent work, Balasubramanian et al.~\cite{balasubramanian2024publicdatasettrackingsocial} collected a dataset from X that highlighted the prominence of keywords such as \emph{Biden}, \emph{Trump}, and \emph{MAGA}, as well as the widespread use of hashtags like \#trump2024, \#maga, and \#bidenharris2024. This dataset reflected public engagement with key candidates and movements and underscored the significant influence of multimedia platforms such as YouTube and X. Moreover, the data illustrated how media outlets, including Fox News and Breitbart, shape political discourse online.

Regarding discourse analysis, several studies have employed various methodologies to explore political polarization and social media rhetoric. For instance, Stefanov et al.~\cite{valence} utilized a combination of supervised and unsupervised learning techniques to detect political polarization on Twitter. Their approach incorporated measures such as valence, graph theory, and contextual embeddings to predict political biases. Similarly, Magno and Almeida~\cite{magno2021measuring} employed word embeddings to explore cultural and social values on a global scale by analyzing large datasets of online communications. Their findings revealed correlations between online sentiment and offline cultural traits, as captured in the World Values Survey.

Hate speech on social media platforms has been extensively explored in various studies. For example, Alkomah and Ma~\cite{hate-speech-1} provided a comprehensive review of textual hate speech detection methods and datasets, while Davidson et al.~\cite{hate-speech-3} focused on automated hate speech detection and the challenges posed by offensive language. Additionally, Saha et al.~\cite{hate-speech-2} proposed novel techniques for hate speech detection that extend beyond traditional methods. In this context, Ottoni et al.~\cite{embd-2} concentrated on right-wing YouTube channels by examining the prevalence of hate speech, violence, and discrimination in both video content and user comments. Their layered methodology, which included lexical analysis, topic modeling, and implicit bias detection, revealed patterns of negative language and discriminatory bias in political discourse.

While most studies on social networks have focused on traditional platforms, there is a noticeable gap in the literature regarding the influence of Discord, particularly within the political context. Although Discord was originally designed as a communication platform for gamers, it has evolved into a multifaceted environment that accommodates a broad spectrum of social and political groups, including far-right organizations. These groups have used Discord to organize events, such as the ``Unite the Right'' rally in Charlottesville, as discussed by Roose~\cite{roose17thiswas}, leveraging its private and community-driven nature for communication and recruitment. Moreover, Heslep and Berge~\cite{heslep2024mapping} demonstrate how hate networks exploit Discord's moderation gaps and third-party tools.

Although numerous studies have investigated political debates on platforms such as Twitter and Telegram, to the best of our knowledge, no comprehensive investigation into political discourse on Discord has been conducted. Given Discord's growing prominence, this study aims to perform an extensive analysis of political discourse on the platform, using the 2024 U.S. presidential election as a case study.

\begin{table*}[H]
\small
\centering
\begin{tabular}{crrrrr} \toprule
\textbf{Alignment} & \textbf{Unique Users} & \textbf{User Messages} & \textbf{Messages/User} & \textbf{Unique Bots} & \textbf{Bot messages} \\ \midrule
\textbf{Democratic}    & 10,149                & 520,614                & 51.29                  & 69                   & 47,123                \\
\textbf{Unaligned} & 71,686                & 31,603,712             & 440.86                 & 618                  & 2,102,070             \\
\textbf{Republican}   & 4,255                 & 555,750                & 130.61                 & 35                   & 76,412                \\
\textbf{Total}   & 83,611                & 32,680,076             & 390.85                 & 675                  & 2,225,605             \\ \bottomrule
\end{tabular}
\caption{Statistics of user and bot posts across server categories. Most of the messages belong to unaligned servers. The unique users and unique bots columns do not sum to total, as some users participate in discussion on multiple server categories.}
\label{tab:statistics}
\end{table*}

\section{Dataset}
\label{sec:dataset}

\subsection{Data Collection}

To analyze political discussions on Discord, we followed the methodology in \cite{singh2024Cross-Platform}, collecting messages from politically-oriented public servers in compliance with Discord's platform policies.

Using Discord's Discovery feature, we employed a web scraper to extract server invitation links, names, and descriptions, focusing on public servers accessible without participation. Invitation links were used to access data via the Discord API. To ensure relevance, we filtered servers using keywords related to the 2024 U.S. elections (e.g., Trump, Kamala, MAGA), as outlined in \cite{balasubramanian2024publicdatasettrackingsocial}. This resulted in 302 server links, further narrowed to 81 English-speaking, politics-focused servers based on their names and descriptions.

Public messages were retrieved from these servers using the Discord API, collecting metadata such as \textit{content}, \textit{user ID}, \textit{username}, \textit{timestamp}, \textit{bot flag}, \textit{mentions}, and \textit{interactions}. Through this process, we gathered \textbf{33,373,229 messages} from \textbf{82,109 users} across \textbf{81 servers}, including \textbf{1,912,750 messages} from \textbf{633 bots}. Data collection occurred between November 13th and 15th, covering messages sent from January 1st to November 12th, just after the 2024 U.S. election.

\subsection{Characterizing the Political Spectrum}
\label{sec:timeline}

A key aspect of our research is distinguishing between Republican- and Democratic-aligned Discord servers. To categorize their political alignment, we relied on server names and self-descriptions, which often include rules, community guidelines, and references to key ideologies or figures. Each server's name and description were manually reviewed based on predefined, objective criteria, focusing on explicit political themes or mentions of prominent figures. This process allowed us to classify servers into three categories, ensuring a systematic and unbiased alignment determination.

\begin{itemize}
    \item \textbf{Republican-aligned}: Servers referencing Republican and right-wing and ideologies, movements, or figures (e.g., MAGA, Conservative, Traditional, Trump).  
    \item \textbf{Democratic-aligned}: Servers mentioning Democratic and left-wing ideologies, movements, or figures (e.g., Progressive, Liberal, Socialist, Biden, Kamala).  
    \item \textbf{Unaligned}: Servers with no defined spectrum and ideologies or opened to general political debate from all orientations.
\end{itemize}

To ensure the reliability and consistency of our classification, three independent reviewers assessed the classification following the specified set of criteria. The inter-rater agreement of their classifications was evaluated using Fleiss' Kappa \cite{fleiss1971measuring}, with a resulting Kappa value of \( 0.8191 \), indicating an almost perfect agreement among the reviewers. Disagreements were resolved by adopting the majority classification, as there were no instances where a server received different classifications from all three reviewers. This process guaranteed the consistency and accuracy of the final categorization.

Through this process, we identified \textbf{7 Republican-aligned servers}, \textbf{9 Democratic-aligned servers}, and \textbf{65 unaligned servers}.

Table \ref{tab:statistics} shows the statistics of the collected data. Notably, while Democratic- and Republican-aligned servers had a comparable number of user messages, users in the latter servers were significantly more active, posting more than double the number of messages per user compared to their Democratic counterparts. 
This suggests that, in our sample, Democratic-aligned servers attract more users, but these users were less engaged in text-based discussions. Additionally, around 10\% of the messages across all server categories were posted by bots. 

\subsection{Temporal Data} 

Throughout this paper, we refer to the election candidates using the names adopted by their respective campaigns: \textit{Kamala}, \textit{Biden}, and \textit{Trump}. To examine how the content of text messages evolves based on the political alignment of servers, we divided the 2024 election year into three periods: \textbf{Biden vs Trump} (January 1 to July 21), \textbf{Kamala vs Trump} (July 21 to September 20), and the \textbf{Voting Period} (after September 20). These periods reflect key phases of the election: the early campaign dominated by Biden and Trump, the shift in dynamics with Kamala Harris replacing Joe Biden as the Democratic candidate, and the final voting stage focused on electoral outcomes and their implications. This segmentation enables an analysis of how discourse responds to pivotal electoral moments.

Figure \ref{fig:line-plot} illustrates the distribution of messages over time, highlighting trends in total messages volume and mentions of each candidate. Prior to Biden's withdrawal on July 21, mentions of Biden and Trump were relatively balanced. However, following Kamala's entry into the race, mentions of Trump surged significantly, a trend further amplified by an assassination attempt on him, solidifying his dominance in the discourse. The only instance where Trump’s mentions were exceeded occurred during the first debate, as concerns about Biden’s age and cognitive abilities temporarily shifted the focus. In the final stages of the election, mentions of all three candidates rose, with Trump’s mentions peaking as he emerged as the victor.

\section{Methods}

In this section, we discuss the methodology used to analyze the characteristics of political discourse on Discord. Political discourse is broadly understood as the communication of ideas, opinions, and debates related to political ideologies, often reflecting the values, beliefs, and actions of different political groups.

\subsection{Analysis of Political Discourse}
\label{sec:txt}

\subsubsection{Political Valence of Messages}

The concept of \textit{political valence} helps to understand how specific words, phrases, and references align with political ideologies. In this analysis, we use it to identify words that are associated with Democratic- and Republican-aligned servers. This metric, introduced in \cite{conover2011political} for hashtags, has been adapted for general terms in subsequent studies~\cite{locatelli2022characterizing}. Unlike topic modeling, which clusters discussions into broad themes without capturing ideological stance, political valence directly quantifies the ideological bias of terms, making it more suitable for analyzing fragmented and short-form political discourse on Discord. By exploring political valence, we aim to uncover the distinct characteristics of the content shared by different groups.

For this work, political valence is defined as:

\[
V(t) = \frac{\frac{N(t, R)}{N(R)} - \frac{N(t, D)}{N(D)}}{\frac{N(t, R)}{N(R)} + \frac{N(t, D)}{N(D)}}
\]

\noindent
where \( N(R) \) and \( N(D) \) denote the total occurrences of all terms in Republican-aligned and Democratic-aligned servers, respectively, while \( N(t, R) \) and \( N(t, D) \) refer to the frequency of a specific term \( t \) in these servers. Additionally, we define \( N(D) = \sum_t N(t,D) \) for a given leaning \( D \). This metric ranges from \([-1, 1]\), where a score of \( -1 \) indicates that a term is exclusively used in Democratic-aligned servers, while a score of \( +1 \) indicates exclusive usage in Republican-aligned servers. A score close to \( 0 \) suggests that the term is used relatively equally across both groups.


We calculate political valence for words and websites shared by Discord users to better understand how political discourse varies between the two opposing groups.

\subsubsection{Embedding Analysis}

\label{sec:embeddings}
A way to understand political discourse on Discord is by examining implicit biases and the semantic contexts of key terms across different ideological spectra. This can be effectively achieved through word embeddings, such as those generated by the \textit{Word2Vec} model\cite{mikolov2013efficient,mikolov2013distributed}. These embeddings map words into a high-dimensional vector space, where semantically similar terms are positioned closer together, capturing contextual relationships in the data.

To investigate these biases, we trained separate \textit{Word2Vec} models for Democratic- and Republican-aligned servers, further segmented by the temporal phases of the election. This approach enables us to analyze how key terms are represented within each context, revealing shifts in their connotations and associations across ideological boundaries and over time. 

However, the dataset segmentation reduces the amount of data available for each model, potentially affecting training effectiveness. To address this limitation, we initialized the training process with a neutral pre-trained \textit{Word2Vec} model. This model was trained on a diverse dataset comprising Wikipedia articles, OpenCrawl data, and movie subtitles \cite{speer2017conceptnet}. 
 
To assess the semantic representation of politically charged terms (e.g., socialism), we adopt the methodology of the Word Embedding Association Test (WEAT)~\cite{caliskan2017semantics}. This approach quantifies the association between target words and reference terms with positive (e.g., good, important) and negative (e.g., bad, terrible) connotations by leveraging the cosine similarity metric in the word embedding space. For a target word \( w \) and two sets of attribute words \( A \) (positive) and \( B \) (negative), the normalized association score \( s(w, A, B) \) is calculated as:

\[
s(w, A, B) = 
\frac{\text{mean} \left( \cos(w, a), \, \forall a \in A \right) - \text{mean} \left( \cos(w, b), \, \forall b \in B \right)}{\text{stddev} \left( \cos(w, x), \, \forall x \in A \cup B \right)},
\]

\noindent
where \( \cos(w, a) \) is the cosine similarity between \( w \) and \( a \). This score quantifies the association of \( w \) with \( A \) and \( B \), identifying implicit biases in word embeddings based on semantic proximity to positive or negative terms. A positive \( s(w, A, B) \) indicates a stronger association of \( w \) with the positive attribute set \( A \), while a negative \( s(w, A, B) \) indicates a stronger association with the negative attribute set \( B \).

In addition to candidate and party names, we arbitrarily selected politically relevant target words, such as \textit{capitalism}, \textit{feminism}, and \textit{market}. For each target word, a set of three positive and three negative reference words was defined based on its context, following the approach outlined in \cite{ottoni2018analyzing}. The selection of positive reference words was informed by previous studies, ensuring coverage of different topics within political discussions \cite{magno2021measuring}. For candidate and party names, a comprehensive set of terms was included to reflect the full spectrum of the debate.

\subsection{Hate Speech}
\label{sec:hatespeech}

The analysis of hate speech within political discourse provides critical insights into the nature and dynamics of discussions, particularly during charged periods such as election campaigns. For this study, we aim to evaluate the prevalence of toxic speech over the course of the electoral cycle and across different political spectra. This approach allows us to understand how toxicity levels fluctuate with key events and how they vary between ideological groups.

In addition, we seek to identify those most targeted by discriminatory discourse in political discussions on Discord. By doing so, we aim to investigate sociopolitical biases in the marginalization of these individuals and examine how such intolerance is intertwined with political debate and the electoral period.

To achieve these objectives, we utilize a RoBERTa-based classification model specifically trained for multi-class hate speech detection \cite{antypas2023robust}. This model is designed to classify text into the following categories: Sexism, Racism, Disability, Sexual Orientation, Religion, Other and Not Hate. The classifier was trained on an extensive dataset of tweets, which closely resembles the Discord environment due to its informal context and the prevalence of short messages. We chose RoBERTa over other options, such as the Perspective API, because of its multi-class classification capability, which enables a more nuanced analysis by covering a broader range of hate speech categories rather than a binary or toxicity-based classification.

\section{Results}
In this section, we present the findings of our study, addressing the research questions (RQs) outlined in the introduction.

To answer RQ1, we investigate user discourse to uncover implicit biases and identify thematic shifts across different political spectra and electoral periods. Specifically, we analyze two aspects of textual content from messages shared on Discord servers: in subsection \ref{sec:valence}, we examine the political valence of specific terms used in discussions, and in the subsection \ref{sec:embedding_results}, we present the embeddings obtained by training \textit{Word2Vec} models on our dataset. Both techniques enable us to explore the ideological leanings of discourse over time.

Furthermore, to address RQ2, we assess the level of toxicity in political discussions in subsection \ref{sec:hate_speech_results}. We identify the primary targets of hate speech and analyze how these dynamics vary across political groups and key electoral moments, using a multi-class speech classifier.

\subsection{Political Valence of Messages}
\label{sec:valence}
\begin{figure}[h]
    \centering
    \includegraphics[width=\linewidth]{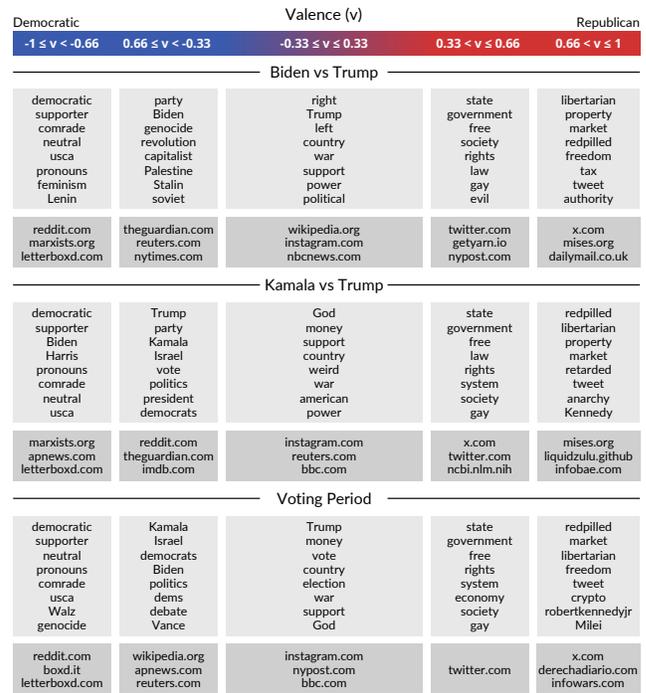}
    \caption{Political valence of top terms and URLs shared and discussed in Democratic and Republican-aligned Discord servers across the studied time period. }
    \label{fig:valenceT1}
\end{figure}

We categorize political valence into five groups, ranging from highly-Democratic to highly-Republican, based on a scale from \( [-1, +1] \).  We opt to make the center interval range [-0.33, 0.33] to account for terms that are used by both sides. Figure \ref{fig:valenceT1} illustrates the top-8 most frequent terms and top-3 most shared URLs within these groups in each period, highlighting discourse shifts during the election year. This analysis excludes links to embedded images, videos, and GIFs, focusing on textual content. We opted to use the most frequent words since we believe that those better represent the subjects commonly discussed on these categories.

\subsubsection{Candidates and Politicians.\\}
\label{sec:politicians} 

In the Democratic side, we observe that throughout all periods the term ``Biden`` is relevant, while Kamala Harris gained more prominence in the discussions only post her presidential campaign announcement, in July 21st, date after Kamala continued to be a top-3 most frequent political term in the moderately-Democratic valence. Also, during the voting period Tim Walz had his first appearance, which is logic since he was then running for the vice-president position.

Even Donald Trump, the main candidate for the Republican Party, appears in the center or slightly Democratic in the valence range depending on the period, likely due to his highly divisive and controversial nature, which sparks discussions on both sides.

Furthermore, the elected vice-president, Vance, appears on the highly-Democratic valence, despite being a Republican. This is intriguing as Vance is often seen as an unpopular choice for vice president when compared to other recent candidates~\cite{vance2024impopular}. It’s noteworthy that the left has seized on this situation to generate discussions about him, while he has largely been ignored on the Republican side. Another interesting appearance in the latest periods is from terms related to Robert Kennedy Jr. in the Republican side of valence, possibly due to him dropping out of the presidential race in support of Trump, and later being picked to be the secretary of the Department of Health and Human Services. 

Lastly, \textit{Milei} shows up as a highly-Republican term during the voting period, referring to Javier Milei, president of Argentina. He is the only non-American politician in the table, and has previously shown support to Trump, being the first foreign leader to meet him since his victory in the elections.

\subsubsection{Political and Economical Ideologies. \\}
\label{sec:ideologies}

\textbf{Democratic Spectrum:\\} Up to the first semester, Democratic-aligned servers focused on Communist and Socialist discourse related terms, such as \textit{comrade}, \textit{Lenin}, \textit{Stalin}, \textit{capitalism} and \textit{revolution} while also sharing websites like \textit{marxists.org}. As the year progressed, however, these discussions and references to Marxist figures and ideals decreased in frequency. 

Discussions about the Israel-Palestine conflict were highly relevant throughout all periods, with terms such as \textit{Israel}, \textit{war}, \textit{genocide}, and \textit{Palestine} appearing in the moderately-Democratic valence. Also, the prevalence of \textit{pronouns} and appearance of \textit{feminism} in highly-Democratic discourse also emphasize an endorsement of equality-related and progressive causes, particularly with respect to the LGBTQIA+ community and women rights.

\textbf{Republican Spectrum:\\}    When examining the Republican side of the discourse, we observe a focus on economical and capitalism related issues, with terms such as \textit{libertarian}, \textit{crypto}, \textit{property}, \textit{market}, \textit{taxes} and \textit{economy}, as well as websites like \textit{mises.org} (an institute focused on economics and libertarianism),  all displaying moderate to highly-Republican valence. One  possible explanation for the prominence of these terms could be  Donald Trump's proposals and general campaign rhetoric, particularly related to taxes, tariffs, and trade, which likely sparked discussions among his supporters. Another notable aspect is that the government-related discussions are more prevalent on the Republican-side, with terms like state, government, and law appearing frequently. 

Additionally,  terms often used as politically incorrect insults, such as \textit{retarded} are among the most commonly used  in the highly-Republican valence, while no insults appear even in the top-100 terms for the highly-Democratic valence.

\subsubsection{News Sources and Associated URLs. \\}
\label{sec:news}

As previously mentioned in \ref{sec:ideologies}, marxists.org and mises.org were commonly shared respectively by the Democratic and the Republican, aligning precisely with left-wing and right-wing ideals.

Democratic-aligned servers also referenced to news outlets such as The Guardian, Reuters, Associated Press and New York Times, with the latter publicly endorsing Kamala Harris campaign~\cite{nyt2024stance}. Meanwhile, Republican-aligned servers were surprisingly associated with two well-known Argentinean media channels, Infobae and DerechaDiario (an auto-declared right-wing news outlet \footnote{In their YouTube channel description 'https://www.youtube.com/c/LaDerechaDiario', they call themselves ``An alternative media to the left-wing hegemony.''}), tying together with Milei's Valence, mentioned before in \ref{sec:politicians}. Another influential source was Infowars which raise concerns about misinformation sharing \footnote{Infowars, known for spreading conspiracy theories and fake news, is concerning due to its influence on political discourse in Republican-leaning communities. https://www.bbc.com/news/world-63243981}. 

Additionally, there was a slight deviation from New York Post links to the Republican side, but it also appears in the balanced Valence during the Voting Period. Other sources such as NBC News, BBC and Wikipedia also stood in the middle-ground of the Valence table, showing no preference from any of the polarized alignments in the Discord platform. Overall, the position of each of the news outlets somewhat reflects the popular perceptions of these sources~\footnote{https://www.allsides.com/media-bias/media-bias-chart}.

\subsubsection{Social Media and Digital Platforms.\\}
\label{sec:social_media}

A few of the URLs found in the table contained hyperlinks to posts on other social media platforms, and their Valence values help us understand the landscape of the preferences from each spectrum.

The first and most important observation is the intriguing positioning of the social media ``X'', previously known as ``Twitter'', as the terms ``x.com'', ``twitter.com'' and ``tweet'' appear heavily related to the Republican. Directly contradicting previous beliefs and studies indicating that Twitter was a predominantly Liberal-leaning platform, with some even saying it affected previous elections by benefiting Democratic candidates \cite{fujiwara2024effect}.

This may be explained by the brand's recent ``rebranding'', which might have attracted a new audience from different political alignments, and is directly corroborated by \cite{balasubramanian2024publicdatasettrackingsocial}, in which is shown that the volume of ``hashtags'' supportive of Republicans, such as \textit{\#MAGA} and \textit{\#Trump} far outweighed the ones related to Democratic backing, as of early 2024.

In contrast, Instagram remains largely non-aligned politically. Its focus on lifestyle, entertainment, and visual content tends to limit its engagement in political discourse, resulting in a more neutral stance in terms of ideological positioning, as backed up by previous studies \cite{allcott24theeffects}. Meanwhile, particularly in the Discord space, Reddit seems to have become more associated with Democrat-aligned communities, as supported by it's Valence values.

\subsection{Embedding Analysis}

\begin{figure}[h]
    \centering
    \includegraphics[width=1\linewidth]{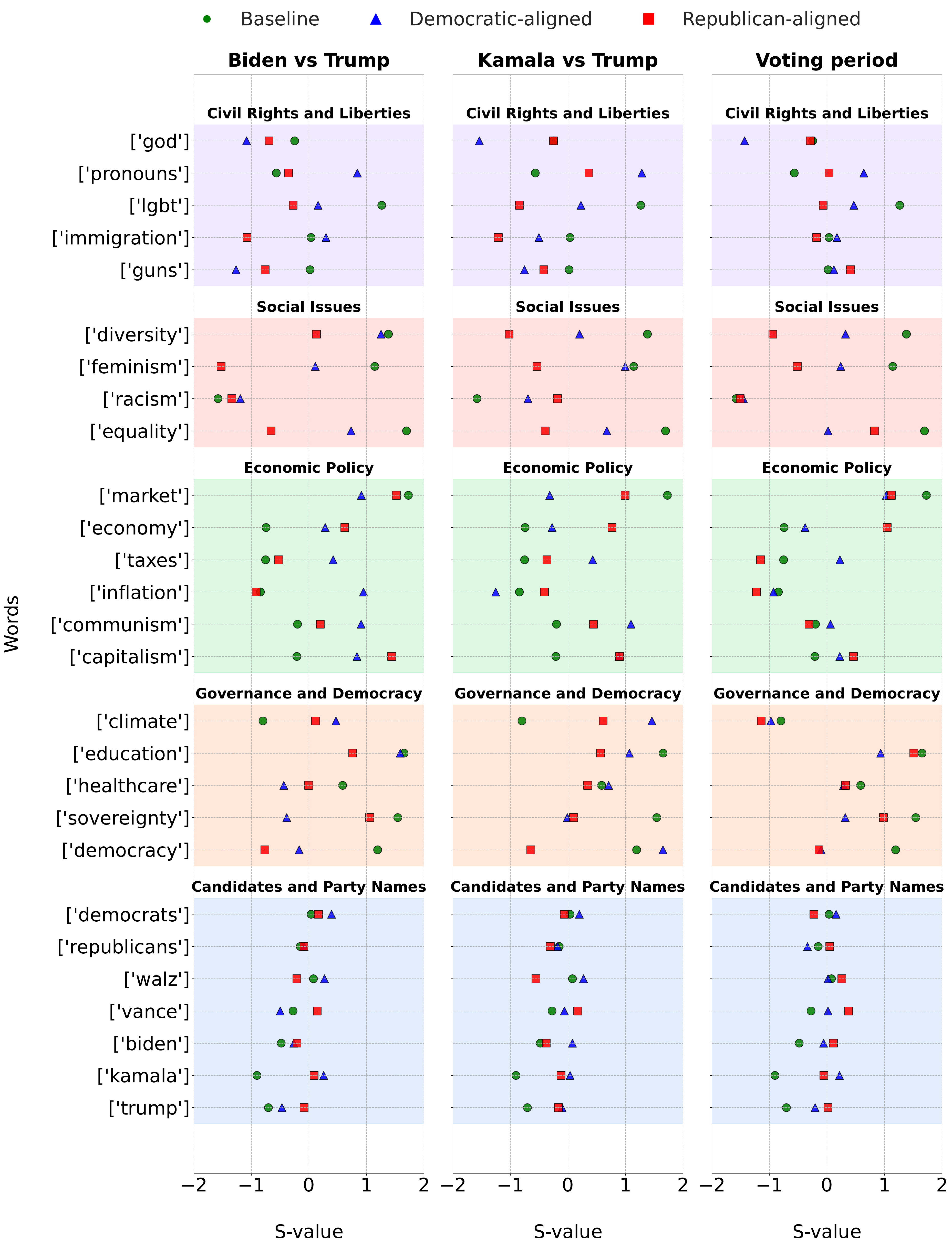}
    \caption{Visualization of S-values computed using models trained on servers with Democratic and Republican alignment for each period of the electoral campaign. The baseline model corresponds to the used pre-trained model. A positive S-value indicates closer alignment with positive terms, while a negative S-value indicates closer alignment with negative terms. }
    \label{fig:weat}
\end{figure}

The Word Embedding Association Test (WEAT) was conducted to uncover implicit biases and semantic associations in the political discourse on Discord servers, seen on Figure \ref{fig:weat}. The purpose of this analysis is to explore how key politically relevant terms are perceived and connected to positive or negative connotations across different ideological contexts and electoral periods.

To ensure comprehensive coverage of political topics, we selected terms spanning a wide range of themes. These terms were organized into five categories: Civil Rights and Liberties, Social Issues, Economic Policy, Governance and Democracy, and Candidates and Party Names. For each term, we defined three positive and three negative reference terms to capture its associations in different contexts. The only exception was the Candidates and Party Names category, where, due to the diverse scenarios and topics these names appear in, we used an extensive list of positive and negative reference terms. This approach ensures a more robust analysis, capturing a wide variety of political themes and contexts.

The first issue concerns Civil Rights and Liberties. Terms related to sexuality and LGBTQIA+ rights appear to be more positively biased on the Democratic side, while being mostly neutral to negative on the Republican side. A notable observation is that the Democratic party, known for its stricter stance on gun laws and more open approach to immigration, has gradually shifted over time, developing a bias more similar to the Republicans on these issues.

New trends emerge in the analysis of Social Issues, bringing several key topics to the forefront, as demonstrated in the hate speech analysis. The term \textit{feminism} stands out, with a predominantly negative bias on Republican-aligned servers. It’s also noteworthy that after Kamala Harris entered the race, both Democratic and Republican servers shifted significantly toward a negative bias on the topic of \textit{diversity}, which correlates with the observed increase in sexist hate speech during this period.

On economic policies, the trends are largely as expected. Words related to \textit{market} or \textit{economy} tend to have a more positive connotation on the Republican side. One notable situation is relating to inflation. During Biden's presidency, many of the critiques against him focused on inflation, which was an issue in late 2022 and early 2023. Interestingly, when Biden was the candidate, the bias towards the term \textit{inflation} in Democratic servers was positive, possibly reflecting that his supporters defend his decisions or the state of the economy. However, once Kamala entered the race, \textit{inflation} shifted to a negative bias, mirroring the Republican perspective.

Governance and Democracy reveal a few notable trends. The term \textit{democracy} saw a sharp increase in positive bias on Democratic servers when Kamala entered the race. This likely reflects her campaign's strong focus on democracy and freedom, which were central to her messaging. Another noteworthy term is \textit{climate}. Until the election, \textit{climate} was largely viewed positively, but as the election approached, its bias turned negative. This shift may be tied to the intensifying discussions around climate change and the evolving political climate.

Finally, we analyze the sentiment dynamics surrounding candidates and party names. In this category, the broader range of terms analyzed resulted in less pronounced differences across the sentiment spectra. However, we can still identify some notable differences. While Joe Biden was the Democratic candidate, both parties exhibited a relatively neutral stance toward Walz. However, upon Walz’s nomination as the vice-presidential candidate, sentiment on the Republican side became markedly more negative.  Interestingly, during the voting period, which coincided with the vice-presidential debate, sentiment on the Republican side shifted back toward a more positive view of Walz. Similarly, Vance experienced a smaller but noticeable shift, where the bias toward him became more positive over time on both the Democratic and Republican sides.

The case of \textit{Biden} presents an intriguing case. Prior to his withdraw of the race, many within his own party members were urging him to step down. As a result, his bias was notably more negative on the left during the first half of the year. However, after he stepped out, his bias became more positive and stabilized toward the middle. Conversely, \textit{Donald Trump} exhibited a similar trend. Even on the Democrat alignment, Trump’s bias became more stable after Biden’s exit, likely due to an increase in his popularity. Interestingly, \textit{Kamala} bias remained relatively stable throughout the entire electoral cycle.

\label{sec:embedding_results}

\subsection{Hate Speech Analysis}
\label{sec:hate_speech_results}

\begin{figure}[h]
    \centering
    \includegraphics[width=1\linewidth]{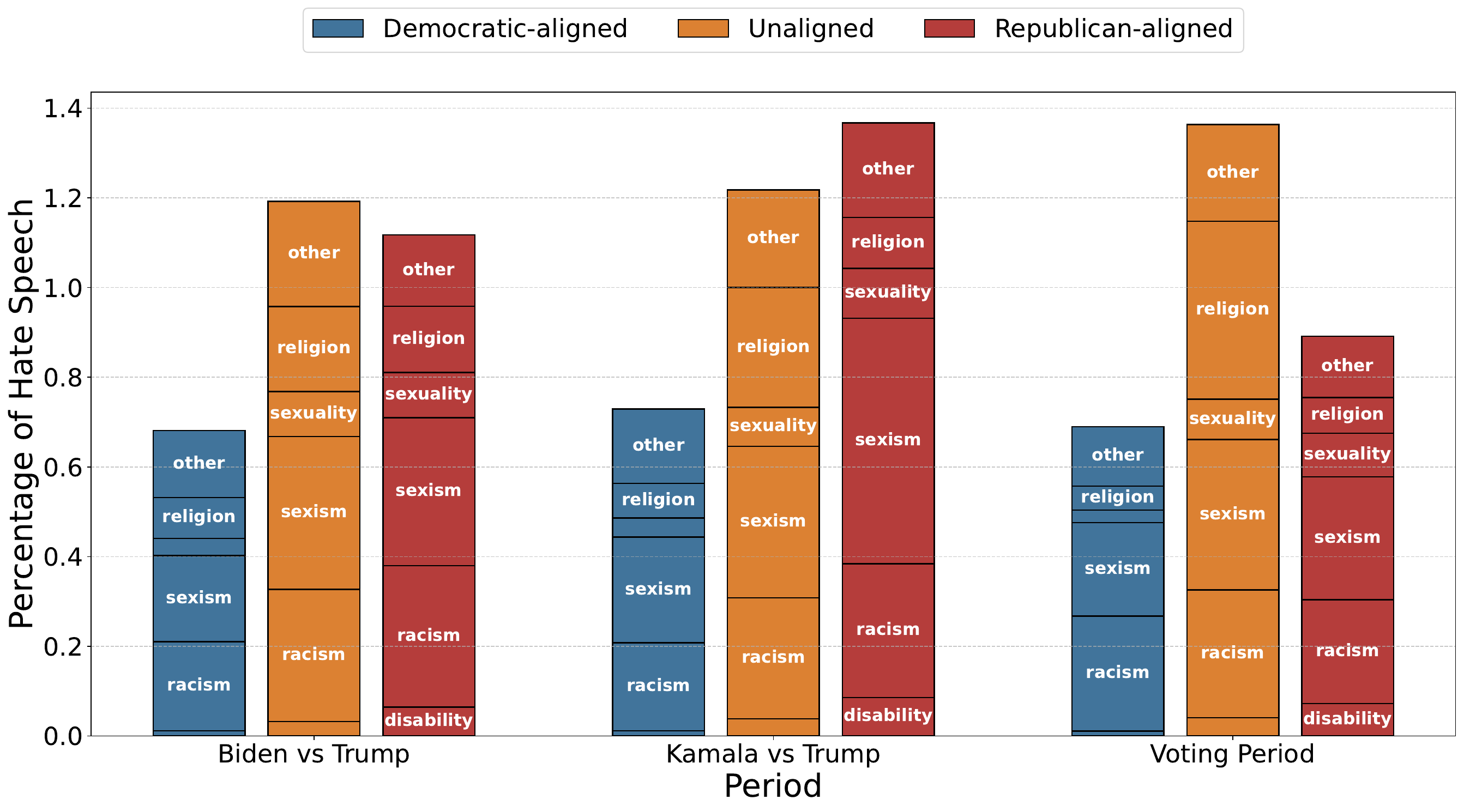}
    \caption{Bar plot of Hate Speech for Democratic Aligned, Republican Aligned and Unaligned in each period. }
    \label{fig:bar-graph-toxic}
\end{figure}

By applying the multi-class hate speech classifier in our dateset, Figure \ref{fig:bar-graph-toxic} displays the average levels of hate speech across Democratic and Republican-aligned and unaligned servers during the periods of interest. The figure provides an overview of the presence of hate speech across multiple classes, offering a general perspective on its distribution. The figure reveals a higher prevalence of hate speech in Republican-aligned and unaligned servers compared to Democratic-aligned servers. 

\begin{figure}[h]
    \centering
    \includegraphics[width=0.82\linewidth]{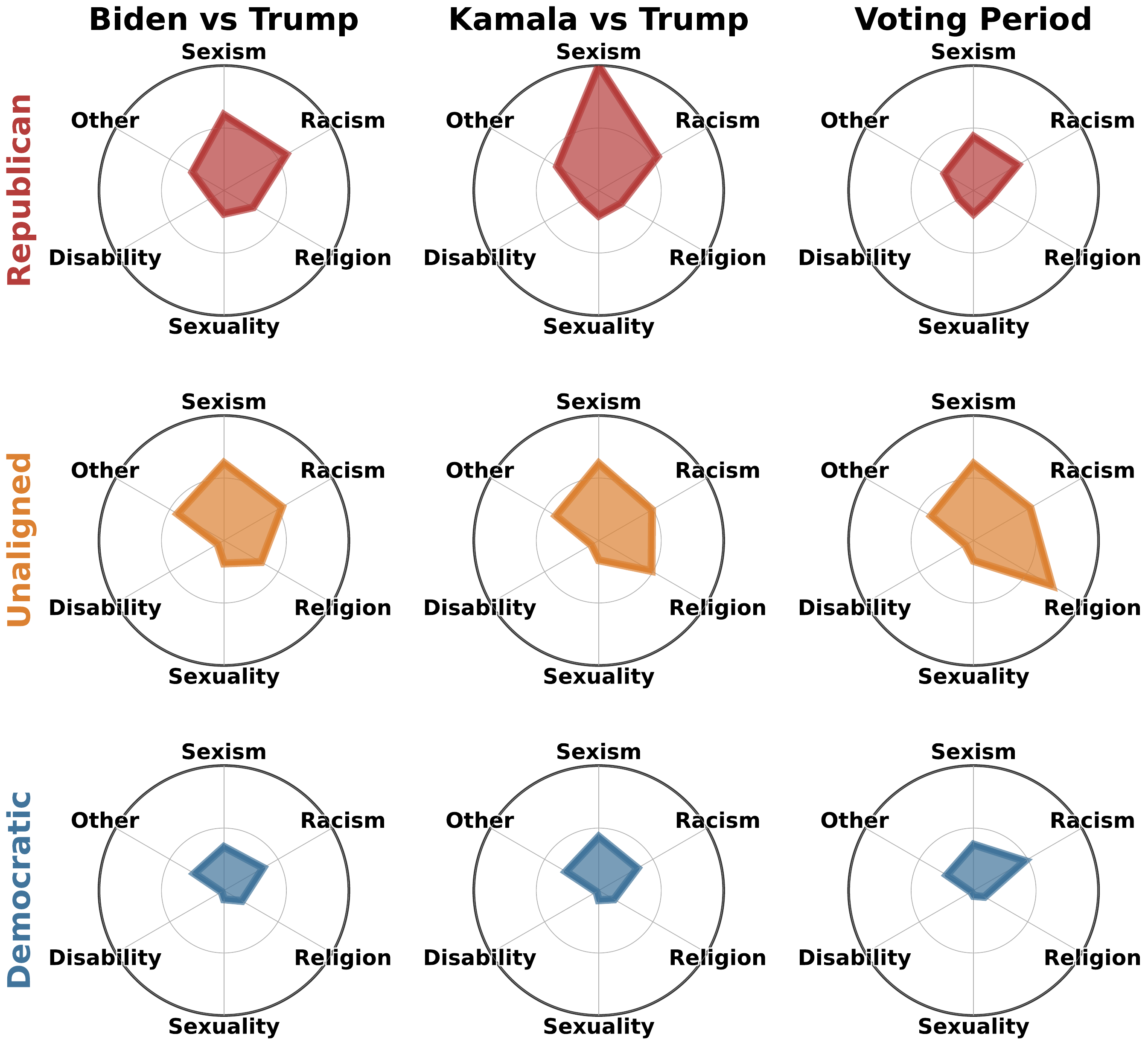}
    \caption{Radar charts displaying the percentage of hate speech messages by period. The highest value, 0.55\%, occurred in the 'Sexism' category during the Kamala vs. Trump race.}
    \label{fig:radar}
\end{figure}

\subsubsection{Hate Speech on Unaligned Servers. \\}

Figures \ref{fig:radar} and \ref{fig:all_hate} provide a detailed analysis of the distribution of hate speech across multiple categories throughout the electoral campaign. The weekly-segmented plot reveal a striking trend in unaligned servers: a consistently high proportion of hate speech across all analyzed categories throughout the observed period. A particularly notable trend is the persistent presence of hate speech in the ``Other'' category, which remains at a significant level over time. Together, these trends, illustrated in Figure \ref{fig:bar-graph-toxic}, demonstrate how all categories cumulatively drive the high global rate of hate speech observed in unaligned servers, especially during the Biden vs. Trump and Voting Periods.

Additionally, it is plausible to hypothesize that unaligned servers, due to their lack of a clear political alignment, function as open forums where diverse opinions and groups converge, helping to explain the consistently high levels of hate speech observed in these servers, unlike explicitly aligned servers (Democratic or Republican), where a greater uniformity of opinions might be expected. Moreover, unaligned servers seems to engage in discussions beyond the scope of the American presidential race. This broader scope is reflected in the significant increase in Religion-related hate speech in early September, as seen in Figure \ref{fig:all_hate}. We hypothesize that this surge is linked to the ongoing Israel-Palestine conflict, which ignited political discussions worldwide, further fueling divisive discourse.

\subsubsection{Hate Speech on Republican and Democratic Servers. \\}

\begin{figure}[h]
    \centering
    \includegraphics[width=1\linewidth]{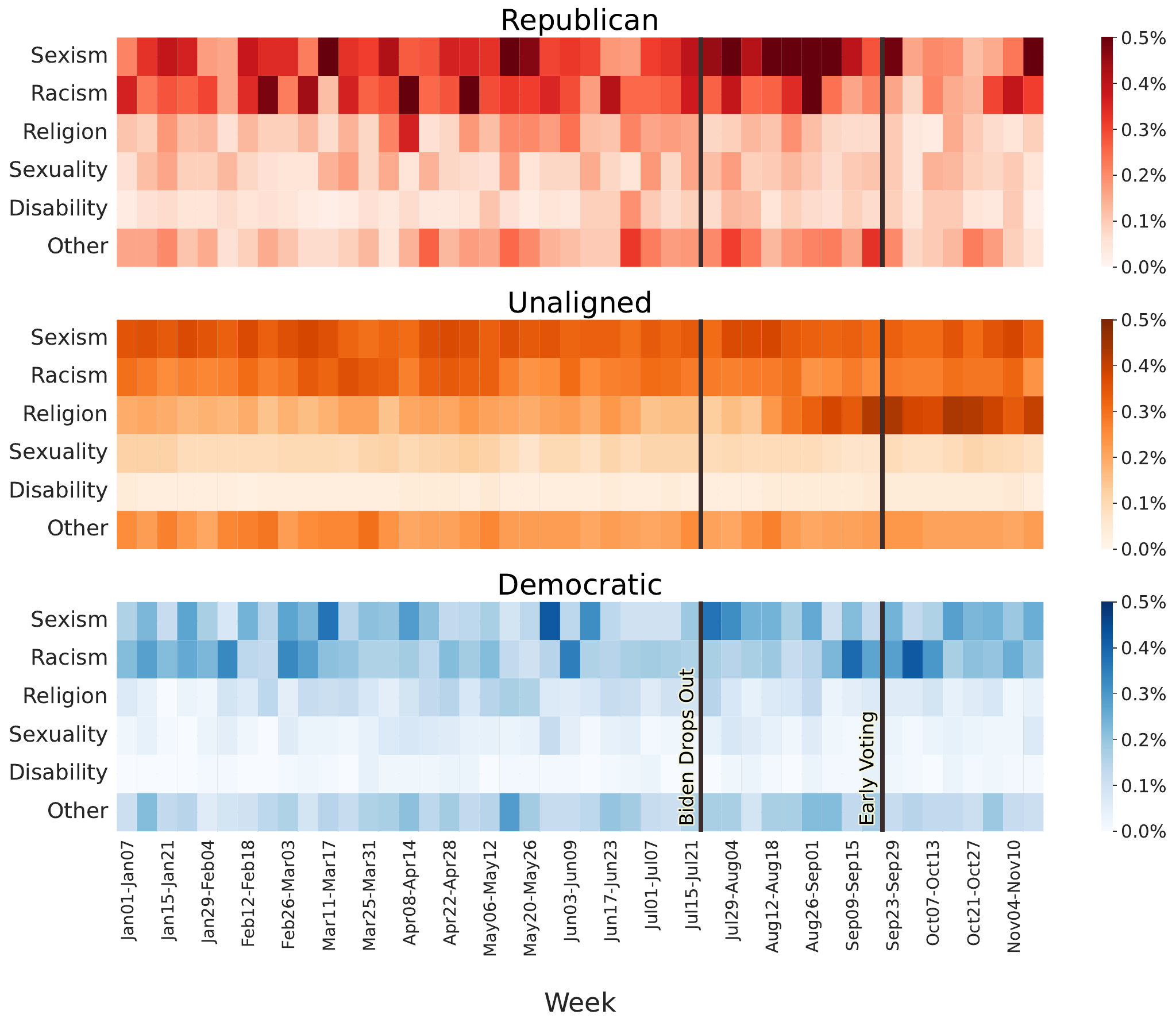}
    \caption{Heatmap of Hate Speech Categories Across Weekly Periods.}
    \label{fig:all_hate}
\end{figure}

On Republican servers, a significant shift in the distribution of hate speech becomes evident following Joe Biden's withdrawal from the campaign and Kamala Harris's entry as the Democratic candidate. As illustrated in Figures~\ref{fig:radar} and \ref{fig:all_hate}, Kamala Harris's candidacy coincides with a notable rise in hate speech targeting sexism, an increase that persists throughout the Kamala vs. Trump period. Notably, a similar, though less pronounced, increase in sexist hate speech is also observed on Democratic servers following her entry.

It is noteworthy that while sexism has risen, racism did not. This disparity is interesting considering that United States has yet to elect a woman as president. Social media platforms often act as mirrors of societal attitudes, amplifying latent sociocultural biases and hate-driven behaviors. These biases, reflected in the volume and nature of hate speech, are closely associated with political narratives and may ultimately impact electoral outcomes. 

Another notable trend in the figure is the significant increase in sexist hate speech observed from the second week of May to the first week of June, evident on both Republican and Democratic servers. We hypothesize that this behavior may be linked to the events of May 16, 2024, during a U.S. House Oversight Committee hearing, where a heated exchange occurred between Representatives Marjorie Taylor Greene (MTG) and Alexandria Ocasio-Cortez (AOC). The confrontation was sparked by a comment from Greene directed at Representative Jasmine Crockett. Greene criticized Crockett's appearance, claiming her false eyelashes hindered her ability to read. Such remarks, when amplified within politically charged groups on social media, may contribute to the rise in online sexism. Discussions in these spaces often devolve into misogynistic attacks and disparaging comments aimed at women in politics \footnote{https://www.theguardian.com/us-news/article/2024/may/17/aoc-v-mtg-house-hearing-chaos}.

Through this analysis, we find that Republican-affiliated servers consistently exhibit higher rates of hate speech related to racism and sexism over the observed period. This trend is clearly illustrated in Figure \ref{fig:bar-graph-toxic}, which also highlight another significant point: the notably low incidence of hate speech related to sexuality in Democrat-affiliated servers. This difference aligns with the historically progressive stance of Democrats, who actively advocate for LGBTQIA+ rights. These findings provide a quantitative basis for explaining the higher overall rates of hate speech in Republican-affiliated servers, driven predominantly by the prevalence of racism and sexism. In contrast, Democrat-affiliated servers demonstrate a stronger commitment to inclusive values, particularly regarding sexuality, further emphasizing the distinction between the two political groups in terms of hate speech dynamics.

\subsubsection{Challenges of Moderation on Discord. \\}

Although Discord has not been extensively studied in academic literature and there is a noticeable gap in analyses of its dynamics, the work \cite{heslep2024mapping} explores the unique challenges of moderation on the platform. They investigate how Discord, in combination with the Disboard website, facilitates the organization of distributed hate networks, exposing significant vulnerabilities in its moderation practices. Unlike more traditional social networks such as Twitter and Facebook \cite{wilson2020hate}, Discord operates on a decentralized model where each community functions autonomously. This decentralization makes centralized moderation extremely limited and creates loopholes for the proliferation of toxic networks. Furthermore, Discord explicitly delegates much of the moderation responsibility to its users, particularly to the administrators and moderators of individual servers \footnote{https://discord.com/safety/360044103531-role-of-administrators-and-moderators-on-discord}.

\section{Conclusion}
\label{sec:concludingremarks}

On this study, we examine political servers on Discord, highlighting consistent differences between Republican- and Democratic-aligned communities throughout 2024. Our findings show that Trump was far more discussed than Kamala Harris, based on the number of mentions. Conversations about Democratic candidates -- Kamala and Biden -- were mostly limited to Democratic-aligned servers, while Trump dominated discussions across all categories of servers, eliciting both positive and negative reactions.  While Kamala’s discourse stayed confined to Democratic-leaning spaces, Trump’s polarizing presence drove high engagement across both sides, engaging  his voter base and contributing to his strong turnout, as presented in Section \ref{sec:politicians}.

Although Discord is primarily used by younger people, who generally lean Democratic \cite{IOP2024}, we found that Republican-aligned servers had higher engagement per user. On average, users in these servers posted twice as many messages as their Democratic-aligned counterparts, resulting in a similar total volume of messages across both categories. Combined with Trump’s significantly higher visibility, this suggests that Republicans were more effective in mobilizing their voter base on Discord.

Our analysis reveals that Republican-aligned servers were significantly more toxic than Democratic-aligned ones, with non political correct insults among the most frequently used terms in those spaces. Some instances of hate speech, including racism and sexism, were notably higher in Republican servers. This trend may have been influenced by Kamala Harris’s identity as a Black woman of Jamaican and Indian descent, making her a target for attacks based on race and gender -- vulnerabilities not as easily exploited against figures like Biden or Trump. Interestingly, while sexist hate speech spiked after Harris officially entered the race, racial hate speech remained stable and even declined slightly over the election cycle. These findings, though preliminary, raise important questions. The U.S. has never elected a woman president, and this pattern might suggest that sexism, often less visible than other forms of bias, could play a larger role than previously understood.

As younger generations increasingly engage in politics, studying the platforms they use to communicate will be critical for understanding modern political movements -- not just in the U.S., but globally. This is essential to addressing challenges like radicalization and polarization that have characterized recent years. This study offers a first look at Discord’s role in contemporary political discourse. Future research should expand on these findings by exploring the broader Discord ecosystem, moving beyond self-identified political leanings and keyword-based filtering to gain deeper insights. 

\subsection{Future Works and Limitations}

One of Discord's distinctive features is the extensive use of bots, future research could focus on investigating the use of bots in the political context on Discord, examining how these automated systems influence discourse, facilitate the spread of political ideologies, or potentially exacerbate the dissemination of hate speech.

Additionally, a more in-depth analysis of the Word Embedding Association Test (WEAT) could be conducted, further exploring the intriguing implicit biases identified in this study. Future work could aim to uncover how these biases are shaped and reinforced within political discussions on Discord. This exploration could provide valuable insights into the mechanisms by which sociopolitical biases emerge and propagate through online interactions.


Despite the large amount of servers and messages collected from public groups using political keywords, it’s important to note that political discussions may also be occurring on public servers that don't explicitly identify with politics or even on private servers. As a result, we cannot fully capture how the platform, as a whole, is engaging with political topics.

Additionally, Discord does not release any user demographic data through its API, such as gender, age, or race -- only usernames are available. This lack of demographic information makes it challenging to fully understand the context of our dataset, as we are unable to analyze the characteristics of the users participating in these discussions.

\section{Ethical Considerations}
\label{sec:ethics}
The data we collected from Discord is accessed via publicly available invitation links, in full compliance with Discord’s terms of service\footnote{https://dis.gd/discord-developer-policy} \footnote{https://discord.com/terms}.  To protect the privacy of users, we employed several anonymization techniques. Specifically, we refrained from identifying individual users or focusing on any hate speech tied to specific users.

\section{Dataset Availability}
The dataset used in this study, which contains messages from public Discord servers during the 2024 U.S. presidential election, is publicly available for further research. It is anonymized to ensure user privacy and organized in a server-specific format, aligned with other relevant literature \cite{aquino2025discordunveiledcomprehensivedataset}, with each server's data stored in an individual JSON file. The dataset is published on Zenodo with DOI: \texttt{10.5281/zenodo.14807501}\footnote{\url{https://zenodo.org/records/14807501}}. For further details, refer to the Discord API documentation\footnote{\url{https://discord.com/developers/docs/reference}}. We encourage researchers to use this dataset for future studies on political discourse, social media dynamics, or related areas.

\begin{acks}
   This work is partially supported by CNPq, CAPES, FAPEMIG, and projects CIIA-Saúde and IAIA--INCT on AI. Virgilio and Wagner hold National Council of Research and Technological Development (CNPq) grants that fund this research (311482/2019-8 and 313017/2022-0, respectively). 
\end{acks}
\bibliographystyle{ACM-Reference-Format}
\bibliography{bibliography}


\end{document}